\documentclass{emulateapj}

\newcommand{\helens}{HE\,1113$-$0641}

\newcommand{\hubble}{\textit{Hubble Space Telescope}}
\newcommand{\hst}{\textit{HST}}

\pdfoutput=1

\begin{document}
\shorttitle{\helens: A New Quadruple Gravitational Lens}
\shortauthors{Blackburne et al.}
\title{\helens: The Smallest Separation Quadruple Lens\\
       Identified by a Ground-based Optical Telescope$^{1,2}$}

\author{Jeffrey~A.~Blackburne\altaffilmark{3}, Lutz~Wisotzki\altaffilmark{4}, \& Paul~L.~Schechter\altaffilmark{3}}

\altaffiltext{1}{Based on observations made with the NASA/ESA Hubble Space Telescope, obtained at the Space Telescope Science Institute, which is operated by the Association of Universities for Research in Astronomy, Inc., under NASA contract NAS 5-26555. These observations are associated with program \#9744.}
\altaffiltext{2}{This paper includes data gathered with the 6.5 meter Magellan Telescopes located at Las Campanas Observatory, Chile.}
\altaffiltext{3}{Massachusetts Institute of Technology, Department of Physics and Kavli Institute for Astrophysics and Space Research, 70 Vassar St., Cambridge, MA 02139; jeffb@space.mit.edu}
\altaffiltext{4}{Astrophysikalisches Institut Potsdam, An der Sternwarte 16, 14482 Potsdam, Germany}
\begin{abstract}

The Hamburg/ESO quasar \helens\ is found to be a quadruple gravitational lens, based on observations with the twin 6.5m Magellan telescopes at the Las Campanas Observatory, and subsequently with the \hubble. The $z_S=1.235$ quasar appears in a cross configuration, with $i'$ band magnitudes ranging from 18.0 to 18.8. With a maximum image separation of $0\farcs{}67$, this is the smallest-separation quadruple ever identified using a ground-based optical telescope. PSF subtraction reveals a faint lensing galaxy. A simple lens model succeeds in predicting the observed positions of the components, but fails to match their observed flux ratios by up to a magnitude. We estimate the redshift of the lensing galaxy to be $z_L \sim 0.7$. Time delay estimates are on the order of a day, suggesting that the flux ratio anomalies are not due to variability of the quasar, but may result from substructure or microlensing in the lens galaxy.

\end{abstract}


\section{Introduction}
\label{sec:intro}

A quasar lensed by a galaxy along the line of sight presents opportunities to study both the gravitational potential that is doing the lensing and the quasar that is lensed.  In particular the inevitable microlensing of the quasar by stars in the lensing galaxies permits estimates of the stellar (as opposed to dark) matter surface density of the lensing galaxy \citep{2004IAUS..220..103S} and of the size of the quasar accretion disk \citep{2007ApJ...661...19P,2007arXiv0707.0003P} and emission line region \citep{2006ApJ...639....1K}.  Quadruply imaged systems are especially useful for such purposes, giving multiple lines of sight through the lens and additional constraints on the lensing potential.

But the rate at which new systems are discovered has leveled off to something like a half dozen per year, with the most recent discoveries coming largely from the Sloan Digital Sky Survey \citep[e.g.][]{2007arXiv0708.0871O}.  Their selection algorithm is incomplete at separations below one arcsecond and worse for quadruple systems than for doubles \citep{2006AJ....132..999O}.  Of the new SDSS lenses, just three are quadruple.  No comparable survey has been carried out in the southern hemisphere.

We report here the discovery of a new quadruply imaged system, \helens. It was observed as part of a program to identify lensed systems among the Hamburg/ESO quasar survey \citep{2000A&A...358...77W}.  While included within the SDSS survey area, it has not, until the present work, been identified as a lensed system.

In \S\ref{sec:obs}, we report the observations made using Magellan and the \hubble. \S\ref{sec:analysis} describes our analysis of the data. In \S\ref{sec:model} we describe a simple model of the lens, and in \S\ref{sec:discuss} we make a rough estimate of the lens redshift. In \S\ref{sec:conclusions} we discuss the conclusions we can come to regarding \helens. Throughout this paper, we adopt a ``concordance'' cosmology with $\Omega_{M} = 0.3$, $\Omega_{\Lambda} = 0.7$, and $h = 0.7$.


\section{Observations}
\label{sec:obs}

\helens\ was originally discovered to be a $z=1.235$ quasar in the Hamburg/ESO digital objective prism survey \citep{2000A&A...358...77W}. Based on its redshift and apparent magnitude $B=17.01$, it was found to have a relatively high lensing probability, and was selected for a follow-up observation.

We observed the object in early 2002 and early 2003 in the Sloan $g'$ and $i'$ bands using the Magellan 6.5 meter telescopes. In Autumn 2003, we observed it in three bands using the ACS and NICMOS detectors on the \hubble. These observations are tabulated in Table \ref{tab:obs}.

\subsection{Magellan 6.5 Meter Imaging}
\label{sec:magobs}

\helens\ was identified as a quadruple gravitational lens on 2002 February 16 using the Baade \mbox{6.5 m} telescope at the Las Campanas Observatory. Seven 60-second exposures in each of the Sloan $i'$ and $g'$ bands, and a single exposure in the $r'$ band, were taken using the Raymond and Beverly Sackler Magellan Instant Camera (MagIC), a 2048$\times$2048 pixel camera with a 2.4 arcminute field of view. The average seeing varied from $0\farcs{}43$ in $i'$ to $0\farcs{}50$ in $r'$ to $0\farcs{}52$ in $g'$. Because of the combination of mediocre seeing with the existence of only a single image in the $r'$ band, and the absence of any corresponding images in the 2003 dataset, we did not carry out any analysis in $r'$.

Second-epoch images were obtained on 2003 January 26, again using MagIC, which had meanwhile been moved to the Clay telescope, 60 meters to the northwest. The three $i'$ band images had an average seeing of $0\farcs{}33$, while the single $g'$ band image had a seeing of $0\farcs{}47$.

The data were bias-corrected, flattened, and combined using standard techniques. The stacked 2003 $i'$ band image may be seen in Figure \ref{fig:big}.

\begin{deluxetable}{lcccc}
\tablewidth{0pt}
\tablecaption{Observations of \helens
\label{tab:obs}}
\tablehead{\colhead{Date} & \colhead{Instrument} & \colhead{Filter} & \colhead{Exposure} & \colhead{Seeing}}
\startdata
2002 February 16 & MagIC   & $g'$ & $7\times60$ sec\phm{0} & $0\farcs{}52$ \\
                 &         & $i'$ & $7\times60$ sec\phm{0} & $0\farcs{}43$ \\
2003 January 26  & MagIC   & $g'$ & $1\times60$ sec\phm{0} & $0\farcs{}47$ \\
                 &         & $i'$ & $3\times120$ sec       & $0\farcs{}33$ \\
2003 November 06 & NICMOS  & $H$  & $3\times640$ sec       & \nodata       \\
                 &         &      & $1\times704$ sec       &               \\
2003 November 07 & ACS/WFC & $V$  & $3\times120$ sec       & \nodata       \\
                 &         &      & $2\times480$ sec       &               \\
                 &         & $I$  & $3\times85$ sec\phm{0} & \nodata       \\
                 &         &      & $2\times346$ sec       &               \\
                 &         &      & $1\times370$ sec       &               \\[-8pt]
\enddata
\end{deluxetable}

\subsection{\hubble\ Imaging}
\label{sec:hstobs}

On 2003 November 6 and 7, \helens\ was observed using both the NIC2 camera of the Near Infrared Camera and Multi Object Spectrometer (NICMOS) and the Wide Field channel of the Advanced Camera for Surveys (ACS). The NICMOS images had 256$\times$256 pixels and a $19\farcs{}2$ field of view, while those from the ACS were significantly larger, with 4096$\times$4096 pixels filling a 3.4 arcminute field of view. Three filters were used, F160W with NICMOS and F555W and F814W with the ACS (hereafter $H$, $V$, and $I$, respectively). Because of the diffraction-limited quality of the images, they were not well sampled, with the width of PSF ranging from 1.5 pixels in $H$ to 2.1 pixels in $I$.

We used the Multidrizzle program of \citet{2002hstc.conf..337K}, version 2.2, to register the ACS images, clean them of cosmic rays, and drizzle them onto a single image per filter. The drizzling process also corrects for geometric distortion arising from the design of the camera. We likewise drizzled the NICMOS images into a single image using the procedure detailed in \citet{hst.dither.handbook}\footnote{The \hst\ Dither Handbook \citep{hst.dither.handbook} is available at:\\*http://www.stsci.edu/hst/wfpc2/documents/dither\_handbook.html}.

The drizzled ACS and NICMOS images of \helens\ may be seen in Figure \ref{fig:images}.


\section{Analysis}
\label{sec:analysis}

The small separation of this lens, combined with the relative faintness of the lensing galaxy, complicated the task of disentangling the four quasar components and the galaxy, particularly for the ground-based data. To address this issue, we fitted for the relative positions and brightnesses of the objects using an iterative process. First we performed a fit to each image for the relative positions of the objects, then averaged the values and repeated the fit, holding constant the relative positions, to determine the photometry.

We used a variant of the DoPHOT photometry package \citep{1993PASP..105.1342S} called Clumpfit to carry out the fits, using empirical PSFs provided by field stars for the quasar components and a pseudo-gaussian profile for the lensing galaxy. (Though this is not a physical profile choice, we found that the choice of galaxy profile had a negligible effect on the goodness of the fit.) We also used DoPHOT to obtain astrometry and aperture photometry for several other stars in the wider-field (ACS and Magellan) images.

\begin{deluxetable*}{ccccccccc}
\tablewidth{0pt}
\tablecaption{Relative Astrometry of \helens\tablenotemark{a}
\label{tab:astro}}
\tablehead{\colhead{} & \multicolumn{2}{c}{B} & \multicolumn{2}{c}{C} & \multicolumn{2}{c}{D} & \multicolumn{2}{c}{G}\\
\colhead{} & \colhead{$\Delta \alpha \cos{(\delta)}$} & \colhead{$\Delta \delta$} & \colhead{$\Delta \alpha \cos{(\delta)}$} & \colhead{$\Delta \delta$} & \colhead{$\Delta \alpha \cos{(\delta)}$} & \colhead{$\Delta \delta$} & \colhead{$\Delta \alpha \cos{(\delta)}$} & \colhead{$\Delta \delta$}}
\startdata
$i'$ (2002) & $-0\farcs{}515$ & $+0\farcs{}428$ & $-0\farcs{}515$ & $-0\farcs{}091$ & $-0\farcs{}148$ & $+0\farcs{}433$ & $-0\farcs{}431$ & $+0\farcs{}188$ \\
$i'$ (2003) & $-0\farcs{}517$ & $+0\farcs{}424$ & $-0\farcs{}523$ & $-0\farcs{}086$ & $-0\farcs{}149$ & $+0\farcs{}432$ & $-0\farcs{}422$ & $+0\farcs{}134$ \\
$V$         & $-0\farcs{}518$ & $+0\farcs{}424$ & $-0\farcs{}523$ & $-0\farcs{}085$ & $-0\farcs{}152$ & $+0\farcs{}427$ &         \nodata &         \nodata \\
$I$         & $-0\farcs{}519$ & $+0\farcs{}422$ & $-0\farcs{}523$ & $-0\farcs{}083$ & $-0\farcs{}152$ & $+0\farcs{}429$ & $-0\farcs{}320$ & $+0\farcs{}145$ \\
$H$         & $-0\farcs{}518$ & $+0\farcs{}425$ & $-0\farcs{}522$ & $-0\farcs{}083$ & $-0\farcs{}146$ & $+0\farcs{}429$ & $-0\farcs{}308$ & $+0\farcs{}169$ \\[-8pt]
\enddata
\tablenotetext{a}{All positions are given relative to component A.}
\end{deluxetable*}

\begin{figure*}
\begin{center}
\includegraphics[width=0.9\textwidth]{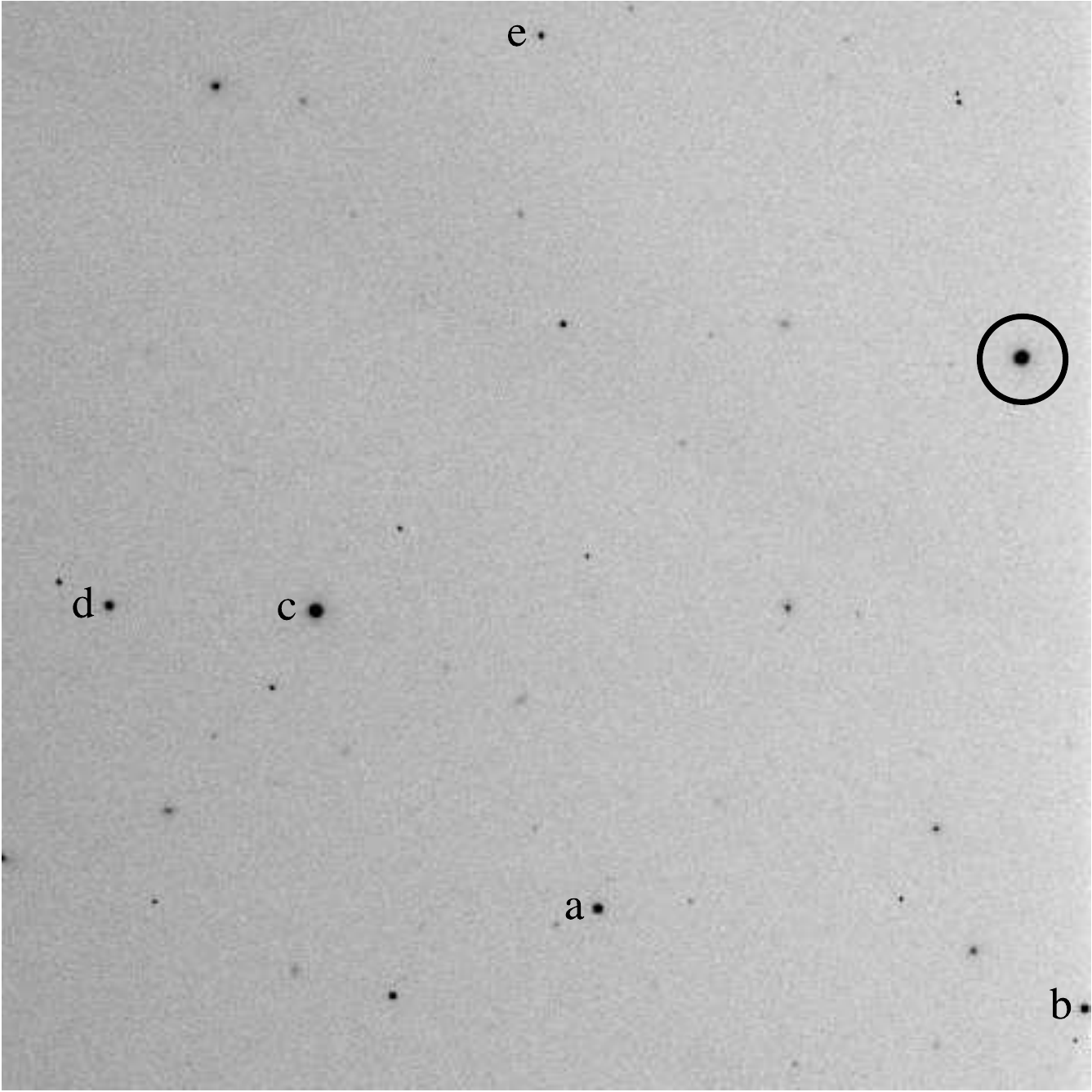}
\end{center}
\caption{2003 $i'$ band image of the \helens\ field, trimmed slightly and binned to $0\farcs{}276$ per pixel. The image is 2.2 arcminutes on a side. The quasar is circled, and the five field stars a through e are labeled. North is up; east is to the left.}
\label{fig:big}
\end{figure*}

\subsection{Magellan data}
\label{sec:maganalysis}

A fit consisting of four empirical PSFs (provided by a field star) was made to the stacked $i'$ band Magellan images. This came to a total of 13 free parameters: two-dimensional position and normalization for all four objects, and the sky level. It was clear from the residual images that a small amount of leftover flux remained; indeed, that the four point sources had been over-subtracted in an attempt to compensate (see Figure \ref{fig:galresid}). So a second fit was attempted using a model with four point sources and a circularly symmetric galaxy; however, there was not enough galaxy flux for the fit to distinguish between radial profiles or converge upon a scale size. We therefore chose a fixed-width pseudo-gaussian profile for the galaxy and repeated the fit, which now had 16 free parameters. The width of the galaxy was set to the width chosen for the ACS data (see \S\ref{sec:acsanalysis}), suitably broadened by the PSF. This fit was able to account for the leftover flux.

The relative astrometry resulting from the 16-parameter fits in $i'$ and other filters is listed in Table \ref{tab:astro}. The relative positions of the four quasar components were then weighted by the inverse of their uncertainties and averaged, yielding final values, which may be found in Table \ref{tab:model}. In the case of the lensing galaxy, only the \hst\ positions were averaged (see \S\ref{sec:acsanalysis}).

Once the relative astrometry had thus been determined, we repeated the fit with fixed relative positions and a fixed galaxy width. The results of this fit may be seen in Table \ref{tab:phot}, and residual images are in Figure \ref{fig:images}.

\begin{deluxetable}{ccc}
\tablewidth{120pt}
\tablecaption{Field Stars a through e\tablenotemark{a}
\label{tab:abcdepos}}
\tablehead{\colhead{} & \colhead{$\Delta \alpha \cos{(\delta)}$} & \colhead{$\Delta \delta$}}
\startdata
a & $+40\farcs{}04$       & $-66\farcs{}50$ \\
b & \phm{0}$-5\farcs{}70$ & $-79\farcs{}16$ \\
c & $+66\farcs{}47$       & $-29\farcs{}91$ \\
d & $+85\farcs{}97$       & $-29\farcs{}07$ \\
e & $+44\farcs{}84$       & $+39\farcs{}90$ \\[-8pt]
\enddata
\tablenotetext{a}{All positions are given relative to component A.}
\end{deluxetable}

For the $g'$ band images, which had poorer seeing, the quasar components were too blurred for a successful fit until relative positions were fixed. The residual images of these fits may also be seen in Figure \ref{fig:images}. There was no indication in the residuals of leftover flux indicative of a lens galaxy, so we conclude that we have not detected it in the $g'$ band.

Aperture photometry was also obtained for several field stars, including those used as model PSFs. The positions of these stars, labeled a through e in Figure \ref{fig:big}, may be found in Table \ref{tab:abcdepos}, and their magnitudes are listed in Table \ref{tab:phot}.

\begin{deluxetable*}{ccccccccccc}
\tablewidth{0pt}
\tabletypesize{\small}
\tablecaption{\helens\ Photometry\tablenotemark{a}
\label{tab:phot}}
\tablehead{\colhead{} & \colhead{$g'$ (2002)} & \colhead{$g'$ (2003)} & \colhead{$i'$ (2002)} & \colhead{$i'$ (2003)} & \colhead{$V$} & \colhead{$I$} & \colhead{$H$} & \colhead{$g'-i'$ (2003)} & \colhead{$V-I$} & \colhead{$I-H$}}
\startdata
A &   18.37 &   18.19 &   18.02 &   17.96 &   18.33 &   18.32 &   18.25 & $+0.23$ & $+0.00$ & $+0.07$ \\
B &   18.28 &   18.24 &   18.09 &   18.02 &   18.40 &   18.35 &   18.27 & $+0.22$ & $+0.05$ & $+0.08$ \\
C &   18.53 &   18.39 &   18.46 &   18.37 &   18.64 &   18.61 &   18.74 & $+0.02$ & $+0.03$ & $-0.13$ \\
D &   18.91 &   18.91 &   18.79 &   18.76 &   19.06 &   19.01 &   18.92 & $+0.15$ & $+0.06$ & $+0.08$ \\
G & \nodata & \nodata &   22.36 &   22.17 & \nodata &   22.47 &   21.05 & \nodata & \nodata & $+1.42$ \\
a\tablenotemark{b} &   20.63 &   20.70 &   18.08 &   18.05 &   20.16 &   17.86 & \nodata & $+2.65$ & $+2.30$ & \nodata \\
b & \nodata & \nodata & \nodata &   18.39 &   20.66 &   18.04 & \nodata & \nodata & $+2.62$ & \nodata \\
c &   19.29 &   19.34 &   16.62 &   16.62 &   18.91 &   16.39 & \nodata & $+2.72$ & $+2.52$ & \nodata \\
d &   18.78 &   18.77 &   18.36 &   18.36 &   18.69 &   18.45 & \nodata & $+0.41$ & $+0.24$ & \nodata \\
e &   22.51 &   22.59 &   20.01 &   20.02 &   22.06 &   19.81 & \nodata & $+2.57$ & $+2.25$ & \nodata \\[-8pt]
\enddata
\tablenotetext{a}{All magnitudes are in the AB photometric system (see \S\ref{sec:maganalysis}).}
\tablenotetext{b}{a through e are field stars.}
\end{deluxetable*}

To enable absolute flux calibration, aperture photometry was obtained for standard stars from the sample of \citet{2002AJ....123.2121S}. For the 2002 data we used PG 1047+003A, and for that of 2003 we used RU 152. We applied a first-order correction for atmospheric extinction when calculating the zeropoints, using extinction coefficients from Table 4 of \citet{2007smith}. It is worth noting that the $u'g'r'i'z'$ system is a broadband approximation to the (monochromatic) AB magnitude system, and is given by
\begin{equation}
m = -2.5\log\frac{ \int d(\log \nu) f_{\nu} S_{\nu} }{ \int d(\log \nu) S_{\nu} } - 48.60 ,
\end{equation}
which is calibrated by synthetic spectra of BD+17 4708. The $u'g'r'i'z'$ deviates from the true AB system by less than 5\%  \citep{2002AJ....123.2121S}, and is presented in Table \ref{tab:phot}.

We estimate the uncertainty in the relative photometry to be 0.1 magnitudes in $g'$ and 0.05 magnitudes in $i'$. Absolute photometry is less certain, with error bars a factor of 1.4 larger. With these uncertainties the data are consistent with a slight overall brightening of all four images between 2002 and 2003, probably caused by intrinsic variability of the quasar. However, they fail to convincingly demonstrate uncorrelated changes in the flux ratios over time, even when combined with \hst\ data; such variations might have been indicative of microlensing.

Despite our use of fixed positions for the $g'$ band images, there are inconsistencies in the $g'-i'$ colors of the quasar components. It is likely that these are due to the confusion caused by mediocre seeing in the $g'$ band.

There can be little doubt that the lensing galaxy has been detected in the $i'$ band in both data sets. However, its size and shape remain poorly constrained. By fitting a fixed pseudo-gaussian to both epochs of data, we were able to estimate its $i'$ band flux, but with substantial uncertainty (0.2 magnitudes of difference between epochs). We were able to determine the position of the galaxy using \hst\ data, but its size and shape remained elusive (see \S\ref{sec:hstanalysis}).

\subsection{\hubble\ data}
\label{sec:hstanalysis}

The \hubble\ images did not suffer from inadequate seeing, but rather from undersampling of the PSF, leading to complications in the interpolation of empirical PSFs. We therefore resampled the ACS images to a scale of $0\farcs{}03$ per pixel, and the NICMOS image to a scale of $0\farcs{}0375$ per pixel, when drizzling.

\subsubsection{ACS}
\label{sec:acsanalysis}

Since the ACS PSF is known to vary across the field of view and also with time, we searched the \hst\ archive for images with a suitable PSF star located close to the position of \helens\ on the chip, and obtained at a time close to 2003 November 7. In the $V$ band, we used a field obtained on 2003 October 7\footnote{The exposure was associated with program \#9756, and started at 4:08 AM.}, and in $I$ we used a field obtained on 2003 November 25\footnote{The exposure was associated with program \#9822, and started at 10:07 PM.}. In order to minimize differences in the PSF caused by the drizzling process, we used the same Multidrizzle process on these images as on the \helens\ images. We chose PSF stars that were close to the correct position on the chip, and were not saturated. In both cases, these stars were about 1 magnitude fainter than image A.

The fits proceeded as they had in the case of the Magellan images. There was appreciable leftover flux in the $I$ band, concentrated in the center of the lens system (see Figure \ref{fig:galresid}). Since the noisy residuals of the quasar components again prevented a measurement of the lensing galaxy's radial profile or scale size, we fit it as a circular pseudo-gaussian with a fixed width of $0\farcs{}35$ (broadened slightly by the ACS PSF). The width was chosen by inspection of the residual image, since the choice had little to no effect on the goodness of fit parameter. No sign of the galaxy was visible in the $V$ band residual image.

\begin{figure*}
\includegraphics[width=\textwidth]{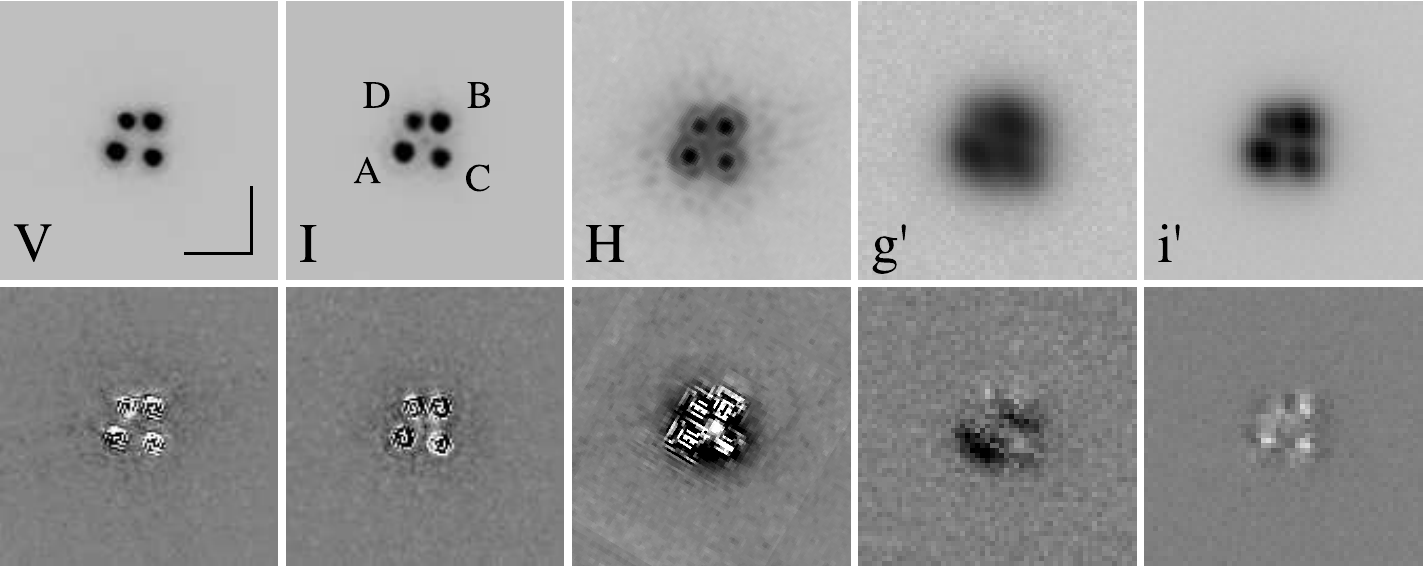}
\caption{Top row: Stacked images of \helens\ from the \hst ($V$, $I$, and $H$) and Magellan ($g'$ and $i'$). Magellan data are from 2003. The images are displayed with logarithmic stretch. Bottom row: Residual images after subtraction of the best model. The images are in a linear stretch from $-20\sigma$ to $20\sigma$, where $\sigma$ is the sky noise. All images are $4\farcs{}0$ on a side.}
\label{fig:images}
\end{figure*}

Both of these fits were repeated once we had determined and fixed the relative positions. The resultant residual images are visible in Figure \ref{fig:images}. The magnitudes were calibrated using AB zeropoint keywords from the \hst\ data headers. The HST broadband flux calibration is based on synthetic spectra of four primary white dwarf stars \citep{1995stsci.conf.bohlin}, and agrees with the AB zeropoints of \citet{2002AJ....123.2121S} to within 3\% \citep{2004AJ....128.3053B}. The photometric data are presented in Table \ref{tab:phot}.

Astrometric measurements were made using DoPHOT on the $I$ band image. A plate solution was found using sixteen USNO-B stars. This solution gives the position of component A as ($11^{\mathrm h}$ $16^{\mathrm m}$ $23\fs{}56$, $-06^{\circ}$ $57'$ $38\farcs{}6$; J2000) to a precision of $0\fs{}01$ in $\alpha$ and $0\farcs{}1$ in $\delta$.

\subsubsection{NICMOS}
\label{sec:nicanalysis}

The analysis of the NICMOS data was similar to that of the ACS data. A 2003 September 5 observation of SA 107-626 provided a model PSF\footnote{The exposure was associated with program \#9875, and started at 4:56 AM.}. In this case, the PSF star was much brighter than the quasar components.

\begin{figure*}
\begin{center}
\includegraphics[width=0.598\textwidth]{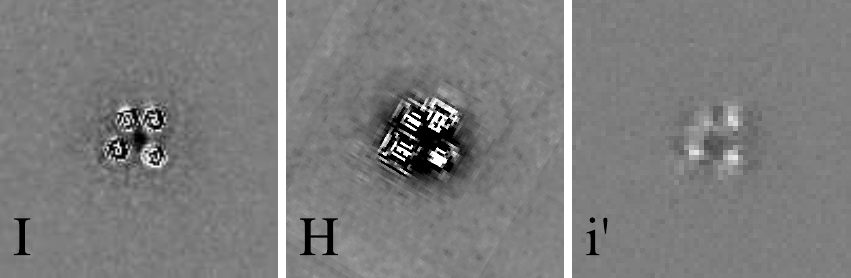}
\end{center}
\caption{Residual images of \helens\ in the redder filters after only four point sources, and no central galaxy, have been fit and subtracted. Leftover flux from the lensing galaxy may be seen near the center. Stretch and size are equal to those of the second row of Figure \ref{fig:images}.}
\label{fig:galresid}
\end{figure*}

Leftover flux from the lensing galaxy was visible in the $H$ band image, as it had been in the $I$ band image (see Figure \ref{fig:galresid}). We again modeled it as a pseudo-gaussian, with the same fixed width (broadened slightly by the NICMOS PSF). 

A second fit was performed with the relative positions of the quasar components fixed to the averaged value. The photometry that resulted from this second fit was calibrated to the AB system using keywords from the data headers, and is visible in Table \ref{tab:phot}. The residual image may be seen in Figure \ref{fig:images}.


\section{Modeling the Lens}
\label{sec:model}

Using the Lensmodel software of \citet{2001astro.ph..2340K}, we modeled the lensing galaxy as a singular isothermal sphere model plus external shear. This model consists of a projected 2-dimensional lensing potential given by
\begin{equation} 
\Psi_{\mathrm{2D}}(\vec{\theta}) = b r - \frac{\gamma}{2} r^2 \cos2(\phi - \phi_{\gamma}),
\end{equation}
where $b$ is the monopole Einstein radius in arcseconds, $r$ and $\phi$ are the radial and angular components on the sky of the position vector $\vec{\theta}$, and $\gamma$ and $\phi_{\gamma}$ are the strength and direction of the external shear. Note that in this convention the position of a companion mass causing a shear would be along the $\pm \phi_{\gamma}$ direction. (No such perturber is observed in this case, consistent with the apparently small shear strength.) This model has seven free parameters, and was constrained by the averaged positions for the four components and the galaxy.

We found the best model to have a radius of $0\farcs{}332$, with $\gamma = 0.04$ and $\phi_{\gamma} = 37.7$ degrees east of north. The source position was predicted to be ($\Delta \alpha \cos{(\delta)}$, $\Delta \delta$)=($-0\farcs{}308$, $0\farcs{}151$) relative to the position of component A. Table \ref{tab:model} contains a summary of the model's predictions, compared to observed data.

\begin{deluxetable*}{cccccccc}
\tablewidth{0pt}
\tablecaption{Astrometry \& Lens Model
\label{tab:model}}
\tablehead{\colhead{} & \multicolumn{2}{c}{Observed\tablenotemark{a}} & \multicolumn{2}{c}{Predicted} & \colhead{Observed\tablenotemark{b}} & \colhead{Predicted} & \colhead{Predicted}\\
           \colhead{} & \colhead{$\Delta \alpha \cos{(\delta)}$} & \colhead{$\Delta \delta$} & \colhead{$\Delta \alpha \cos{(\delta)}$} & \colhead{$\Delta \delta$} & \colhead{$\mu$} & \colhead{$\mu$} & \colhead{Time Delays\tablenotemark{c}}}
\startdata
A &           $0''$ &           $0''$ & $-0\farcs{}002$ & $+0\farcs{}001$ &       $+$16.0 &       $+$12.6 &         0  \\
B & $-0\farcs{}518$ & $+0\farcs{}424$ & $-0\farcs{}518$ & $+0\farcs{}423$ &       $+$15.7 &       $+$15.7 & \phm{0}7.9 \\
C & $-0\farcs{}522$ & $-0\farcs{}085$ & $-0\farcs{}524$ & $-0\farcs{}084$ &       $-$10.2 & \phm{0}$-$9.6 &       36.9 \\
D & $-0\farcs{}150$ & $+0\farcs{}429$ & $-0\farcs{}150$ & $+0\farcs{}428$ & \phm{0}$-$8.6 &       $-$16.7 &       16.4 \\
G & $-0\farcs{}313$ & $+0\farcs{}158$ & $-0\farcs{}314$ & $+0\farcs{}148$ &       \nodata &       \nodata &    \nodata \\[-8pt]
\enddata
\tablenotetext{a}{Weighted average of positions from $i'$ band data and HST data.}
\tablenotetext{b}{Flux ratios from H band data; normalized so that component B matches the model. Negative magnification denotes saddle point images.}
\tablenotetext{c}{In hours.}
\end{deluxetable*}

The model fits the component positions very well, even with the tight constraints provided by the \hst. It does not, however, correctly predict the flux ratios. This is in keeping with experience; optical flux anomalies can be seen in many lensed quasars \citep{2003ApJ...598..138K}. In particular, the D component, a saddle point image, is predicted to be brightest, but is observed to the be the faintest, too faint by as much as a factor of 2.5. The predicted position of the lensing galaxy is $0\farcs{}01$ from the observed position. This is well within the estimated measurement error.

Finally, the model allows us to predict the time delays between the images, given a lens redshift. We used $z_L=0.7$, as estimated in \S\ref{sec:discuss}. This yields the predicted time delays seen in Table \ref{tab:model}, with the maximum delay being $\sim$1.5 days. We also calculated predicted time delays for $z_L=0.4$ and $z_L=1.0$; these changes reduced and increased (respectively) the time delays by a factor of $\sim$3. This strong dependence suggests that a measurement of the time delays might constrain the lens redshift; however, the unknown radial mass profile of the lensing galaxy is likely to have a similarly strong effect on time delays.

By way of comparison, HE\,0435$-$1223, which has a shape similar to that of \helens\ but a larger image separation, has a maximum time delay of two weeks \citep{2006ApJ...640...47K}.


\section{Estimating the Lens Redshift}
\label{sec:discuss}

In order to estimate the redshift of the lensing galaxy and determine what its $I-H$ color could tell us about its morphology, we combined the results of our lens model with properties of typical galaxies.

From the lens strength $b=0\farcs{}332$ we found the line-of-sight velocity dispersion of the lens using
\begin{equation}
b=\frac{D_{LS}}{D_{S}}\frac{4 \pi \sigma^2}{c^2}
\end{equation}
\citep{1996astro.ph..6001N}, where $D_{LS}$ is the angular diameter distance from the lens to the source, and $D_{S}$ is the angular diameter distance from the observer to the source. These distances depend on both the source redshift $z_S=1.235$ and the unknown lens redshift $z_L$.

By combining this equation with the Faber-Jackson relation \citep{1976ApJ...204..668F} for elliptical galaxies, or the Tully-Fisher relation \citep{1977A&A....54..661T} for spiral galaxies, we generated a predicted observed magnitude for each filter as a function of lens redshift. The Faber-Jackson relation is given by
\begin{displaymath}
M^0_T(B) = -19.4 + 5 \log{h} - 10(\log{\sigma}-2.3)
\end{displaymath}
\citep{1982ApJ...256..346D}, which becomes
\begin{equation}
M_B = -18.9 + 5 \log{h} - 10(\log{\sigma}-2.3),
\end{equation}
after applying the extinction correction $B_T - B^0_T = 0.22$ \citep{1976RC2...C......0D}, and with $B_J = B_T + 0.29$ \citep{1986MNRAS.221..233P}. The Tully-Fisher relation is the same, but with $\sigma$ replaced by circular velocity, which for an isothermal sphere is just $\sqrt{2}\sigma$.

The predicted magnitude in the $i'$ band is given by
\begin{equation}
m_{i'}=M_B+DM(z_L)+K_{B,i'}(z_L)
\end{equation}
where $DM(z_L)$ is the cosmological distance modulus, and $K_{B,i'}(z_L)$ is the generalized K-correction between the lensing galaxy's rest-frame $B$ band magnitude and the observed $i'$ band magnitude \citep[see, e.g.,][]{2002astro.ph.10394H}. To calculate the K-correction for an elliptical galaxy at each potential lens redshift, we used an SED generated by the \citet{2003MNRAS.344.1000B} spectral evolution code. Our model consisted of a solar-metallicity, instantaneous starburst at a redshift of 3.0, followed by passive evolution. For a spiral galaxy, we used an empirical Scd galaxy spectrum from \citet{1980ApJS...43..393C}, redshifted appropriately.

The observed magnitudes of the lensing galaxy matches those predicted for an elliptical galaxy for a range of redshifts $0.4 \lesssim z \lesssim 1.0$. A spiral galaxy model also matches the $H$ band observations, but would be brighter than observed at all redshifts in $I$ by $\gtrsim 1$ magnitude. The galaxy's brightness and colors seem to be more consistent with an elliptical galaxy than a spiral.

In addition to this method, we estimated the probability distribution of the lensing galaxy's redshift by calculating lensing optical depth as a function of redshift, following the approach of \citet{1992ApJ...384....1K}. We found that the median redshift was 0.66, with a 68\% confidence interval of [0.41,0.88]. This is consistent with the results of the Faber-Jackson method.

\section{Conclusions}
\label{sec:conclusions}

The $z_S=1.235$ quasar \helens\ is lensed into a cross configuration, with four components ranging from 18.0 to 18.8 magnitudes in $i'$. The maximum image separation is $0\farcs{}67$. A combination of ground-based and \hst\ imaging has yielded reliable astrometry and photometry of the four quasar components of \helens, as well as a good estimate of the position of the lensing galaxy. However, we were unable to measure the size or morphology of the galaxy. By assuming a circular pseudo-gaussian shape and fixing a width for the galaxy, we were able to estimate its flux in the redder bands.

A singular isothermal sphere (SIS) model succeeded in matching the positions of the quasar components and of the lensing galaxy, but was unable to match the observed flux ratios. Based on what has been seen with other gravitational lenses, it seems likely that this is due to perturbations from stellar microlensing or dark matter substructure in the lens galaxy.

The redshift of the lens galaxy remains elusive, but we estimate that $z_L \sim 0.7 \pm 0.3$, based on its observed flux and colors. At this redshift, its velocity dispersion, as measured by its lensing potential, would be 180 km/s.

Because of its small separation, \helens\ will likely prove difficult to monitor using ground-based telescopes, and hope for measuring a spectroscopic redshift of the lensing galaxy is faint. Nevertheless, it is an interesting example of a small-separation lens, and may prove useful for studies that can take advantage of telescopes with very good seeing.

\acknowledgements 

Support for program \#9744 was provided by NASA through a grant from the Space Telescope Science Institute, which is operated by the Association of Universities for Research in Astronomy, Inc., under NASA contract NAS~5-26555. J.~A.~B.\ and P.~L.~S.\ gratefully acknowledge support from NSF Grant AST-0206010.

\end{document}